# Predicting Gene Expression Between Species with Neural Networks

Peter Eastman[1] and Vijay S. Pande[1]

[1]Department of Bioengineering, Stanford University, Stanford, CA 94305

## Abstract

We train a neural network to predict human gene expression levels based on experimental data for rat cells. The network is trained with paired human/rat samples from the Open TG-GATES database, where paired samples were treated with the same compound at the same dose. When evaluated on a test set of held out compounds, the network successfully predicts human expression levels. On the majority of the test compounds, the list of differentially expressed genes determined from predicted expression levels agrees well with the list of differentially expressed genes determined from actual human experimental data.

## Introduction

An important problem in biomedical research is to predict the result of an experiment on one species based on experiments conducted on a different species. This is especially vital in preclinical pharmaceutical research. Due to the cost, ethical issues, and time requirements of testing drug candidates on humans, they are first tested on animals and only advanced to clinical trials if the results are sufficiently promising. Unfortunately, laboratory animals are often a poor proxy for humans [1]. For example, a drug that appears safe in rodents may turn out to be highly toxic in humans [2], and vice versa. One cannot assume humans will respond in exactly the same way as test animals. Instead, we view the data collected from animal experiments as raw data, then seek to predict the result of conducting a similar experiment on humans, taking the many biological differences into account.

A closely related problem is to predict the result of *in vivo* experiments based on data from *in vitro* experiments with cell cultures or organoids [3]. A human cell line is very different from a live human, and its response to a chemical may be very different. The goal is to use the data from *in vitro* experiments to predict how a human would respond, taking all of the biological differences into account.

In this study, we consider the particular problem of trying to predict gene expression levels. For example, given expression data for a rat treated with a particular drug, predict how that same drug would affect expression levels in a human. Software for doing this sort of prediction has existed for some years [4], but it has generally worked in a very primitive way based only on homology. For every gene in the target organism, it looks for an ortholog in the source organism then assumes the two genes would have identical expression levels. This takes none of the biological differences between organisms into account.

A more recent study tried to improve on this by filtering the list of orthologs to only those displaying a consistent, monotonic dose response in both species [5]. The results show that translating expression between species is possible, but also demonstrate the limited power of this approach. The predictive genes varied between compounds, and for some compounds no predictive genes at all were identified. Simply copying expression values between orthologs is clearly not a reliable technique.

In this study we try a different approach, using machine learning to train a neural network that can translate full expression profiles between species. We use expression data from the Open TG-GATES database [6] for rat and human liver cells treated with various compounds. Samples are paired that share the same compound, dose, and time point. Approximately 90% of the data (1024 rat/human sample pairs for 125 compounds) is used to train a neural network that takes rat expression levels as input and produces human expression levels as output. The remainder of the data (98 rat/human sample pairs for 14 other compounds) is set aside as a test set. The result is a model that can take arbitrary rat expression profiles as input and predict a complete expression profile for human cells treated with the same compound at the same dose.

This work should be viewed as a proof of concept. Ideally we would like to predict the effect of a drug *in vivo*, but because Open TG-GATES only provides *in vitro* data for humans, we were unable to test that. In principle the model and training procedure described here should be equally applicable to any type of expression data, whether for cell lines, organoids, or living organisms. One simply needs a sufficient set of paired experimental samples to train it on.

## Results

### Prediction Accuracy

We can evaluate the accuracy of the model by feeding rat expression levels into it, then comparing the predicted human expression levels $y_{pred}$ to those which were actually measured in the corresponding experiment on human cells $y_{true}$. Figure 1 shows a scatterplot of the predicted vs. real expression levels for a typical sample in the test set. There is a strong linear correlation between the two, showing the model has successfully learned something. Over the entire test set (a total of 1,994,496 measurements), the real and predicted expression levels have a correlation coefficient of 0.697 and a mean absolute error (MAE) of 0.158. There is considerable variation in accuracy between samples. Over the 98 samples in the test set, the correlation coefficient ranges from 0.104 to 0.929, with a median of 0.791. The MAE ranges from 0.098 to 0.320, with a median of 0.139. We will return later to the question of why the predictions are more accurate for some samples than others.

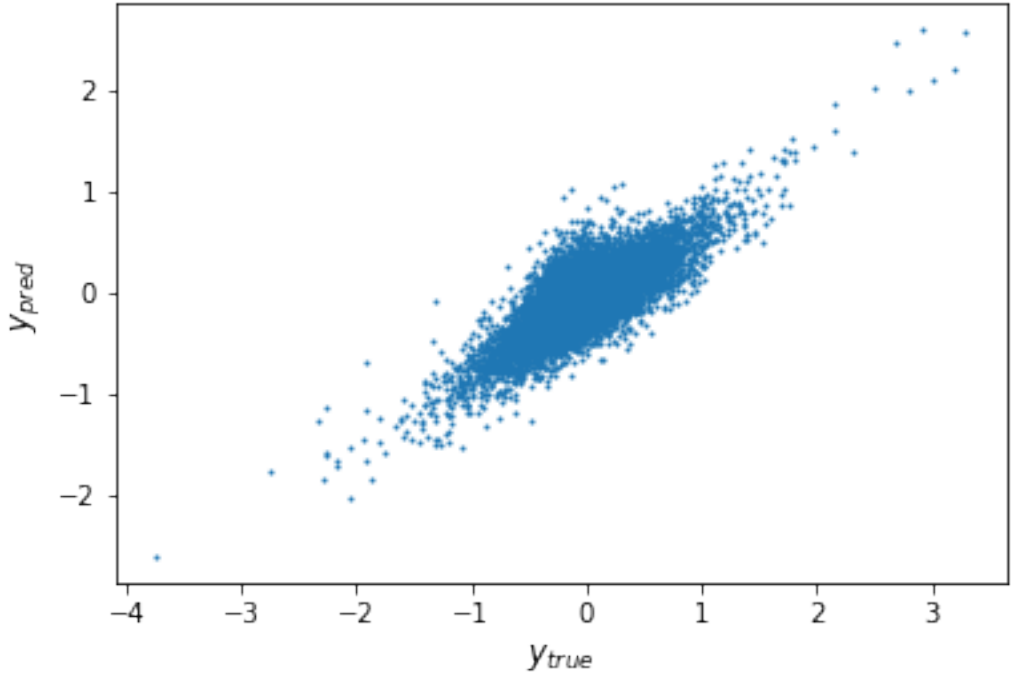

Figure 1. Predicted vs. experimental expression levels for all human genes in a representative sample from the test set.

Ideally one would like to know in advance which values are most accurate. In addition to predicting the expression level of every human gene, the neural network is also trained to estimate the standard deviation in each of its output values. Assuming these estimates are accurate, this allows a user to assess the reliability of conclusions drawn from the predicted expression levels. Figure 2 shows a scatterplot and heat map of the true error $y_{pred}$-$y_{true}$ versus the predicted standard deviation $\sigma_{pred}$ for every gene in a typical test sample. There is a clear connection between the two, with the error magnitude growing with $\sigma_{pred}$. For genes that are predicted to have a small standard deviation (those at the left edge of the scatterplot), the true error is consistently small.

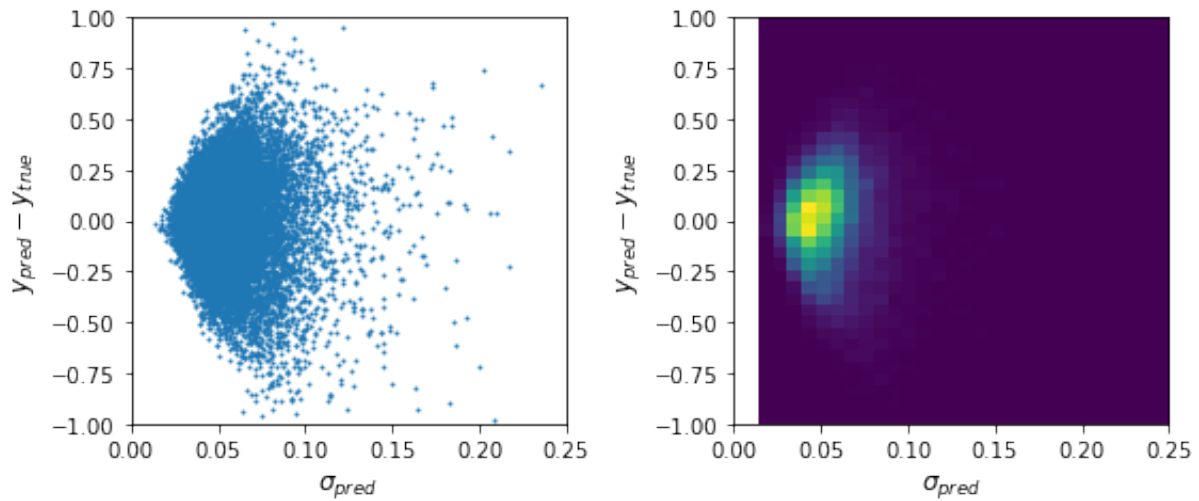

Figure 2. True error versus predicted standard deviation for all human genes in a representative sample from the test set. The heat map (right) emphasizes that most genes fall in the low-error region of the plot.

Figure 3 shows a histogram of $|y_{pred} - y_{true}|/\sigma_{pred}$, the absolute error measured in units of the predicted standard deviation, over all data points in the test set. Most predicted expression levels are within a few standard deviations of the true value, although the distribution clearly decays more slowly than a normal distribution. This again demonstrates that the predicted standard deviations provide useful information about the accuracy of predicted expression levels.

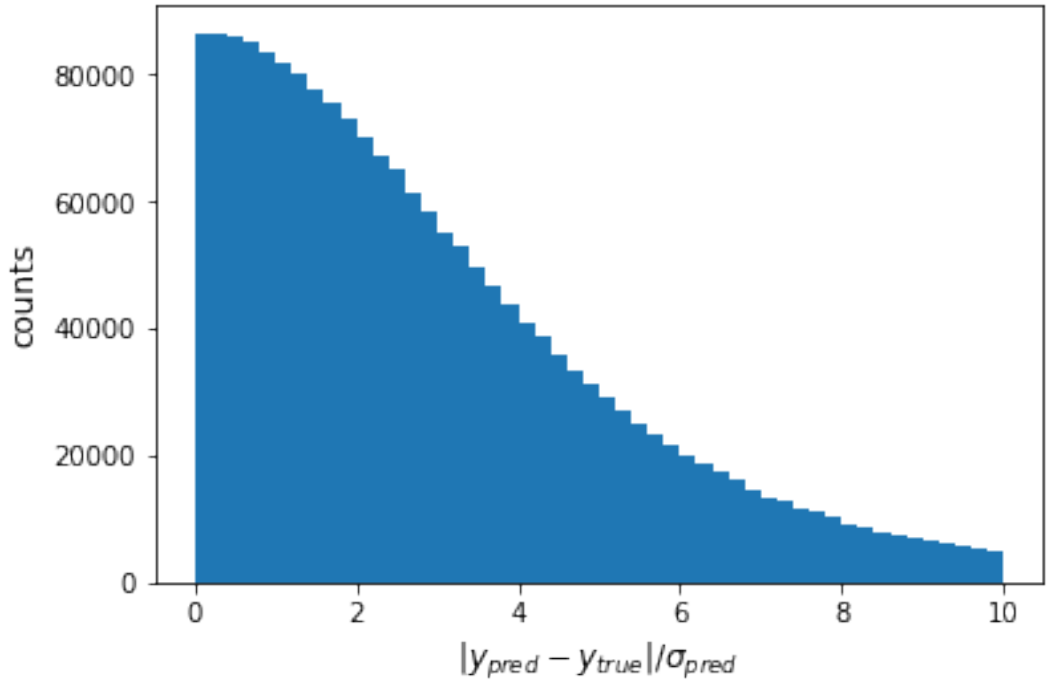

Figure 3. Histogram of the absolute error in expression values, measured in units of the predicted standard deviation.

Now consider the variation in error between samples. Figure 4 shows a scatterplot of MAE versus mean predicted standard deviation for the 98 samples in the test set. There is some correlation between the two, but it is fairly weak (correlation coefficient 0.17). The predicted uncertainties appear to be less useful for judging the relative accuracies of samples than they are for judging the relative accuracies of genes within a single sample.

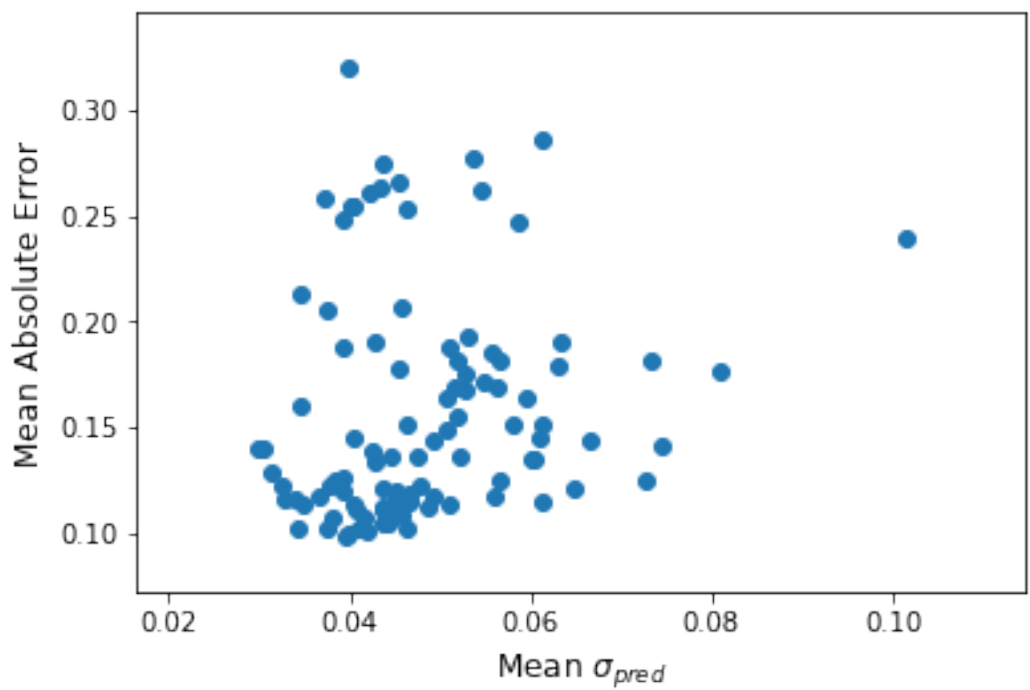

Figure 4. Mean absolute error versus mean predicted standard deviation for the 98 samples in the test set. Means are taken over all the genes for which we have expression data.

In summary, the predicted uncertainties are a useful tool that provides insight into the likely range of each value, but they must be interpreted cautiously. One should be especially careful about using them in statistical tests as if they were true standard deviations. They are only estimates and cannot form the basis of a rigorous statistical analysis.

**Differentially Expressed Genes**

For most practical purposes, the ultimate test of predicted expression levels is whether they can be used to identify differentially expressed genes. We evaluate this by computing a list of the 100 most highly differentially expressed genes for each of the 14 compounds in the test set. For each one we compare the earliest time point (2 or 8 hours) control sample to the latest time point (24 hours), highest dose sample. Genes are ranked based on the absolute difference in their normalized log scale expression levels between these two samples. This analysis is performed twice: once for the experimental data on human cells, and once for the predicted human expression levels based on experimental rat data.

This is, of course, a very simple way of identifying differentially expressed genes. In practice one would typically use a more sophisticated statistical analysis that takes into account all dose levels and time points. For our present purposes, we just need a simple analysis that can be applied in a consistent way for all compounds. We seek a qualitative answer to the question of whether the model's predictions provide useful information about which human genes are most affected by each compound.

Table 1 compares the two lists of genes for each compound (one based on true data and one based on predicted data) and shows how many genes they have in common. The number varies from only 6 of 100 genes for papaverine, up to 85 of 100 genes for tamoxifen. For 8 of the 14 test compounds, the two lists share the majority of their genes in common. Note that if the genes were chosen completely randomly, we would expect the average number of common genes to be less than one. Having even 6 genes in common therefore indicates that the model is producing significant information about the true list of genes, although it might not be enough information to be useful for a particular purpose.

| Compound | # Genes |
|---|---|
| acarbose | 63 |
| aspirin | 66 |
| chloramphenicol | 59 |
| danazol | 13 |
| diltiazem | 59 |
| famotidine | 69 |
| furosemide | 15 |
| hydroxyzine | 53 |
| imipramine | 55 |
| methapyrilene | 7 |
| papaverine | 6 |
| promethazine | 36 |
| tamoxifen | 85 |
| tunicamycin | 12 |

Table 1. The number of the top 100 differentially expressed genes in common between the experimental and predicted expression levels.

We now consider why the results are more accurate for some compounds than others. Generally speaking, any machine learning model will tend to be most accurate when used on data similar to the data it was trained on. This suggests two hypotheses for why our neural network produces more accurate results for some compounds than others.

Hypothesis 1. Some compounds affect expression levels in ways that are very different from those in the training set, leading to a reduction in accuracy on those compounds. If this hypothesis is correct, we should expect to see a dose effect. The control samples for those compounds should have no more error than any other sample, but the error should increase with dose.

Hypothesis 2. There may have been differences in experimental conditions unrelated to the particular compound being tested. The Open TG-GATES data was generated by multiple organizations over a period of ten years. Efforts were made to keep experimental conditions consistent, but some variation was inevitable. If this hypothesis is correct, we should expect to see unusually large errors in the control samples for some compounds, even though all control samples should in principle be equivalent.

Figure 5 shows how the MAE varies with both compound and dose. We see that the compounds divide into a few categories.

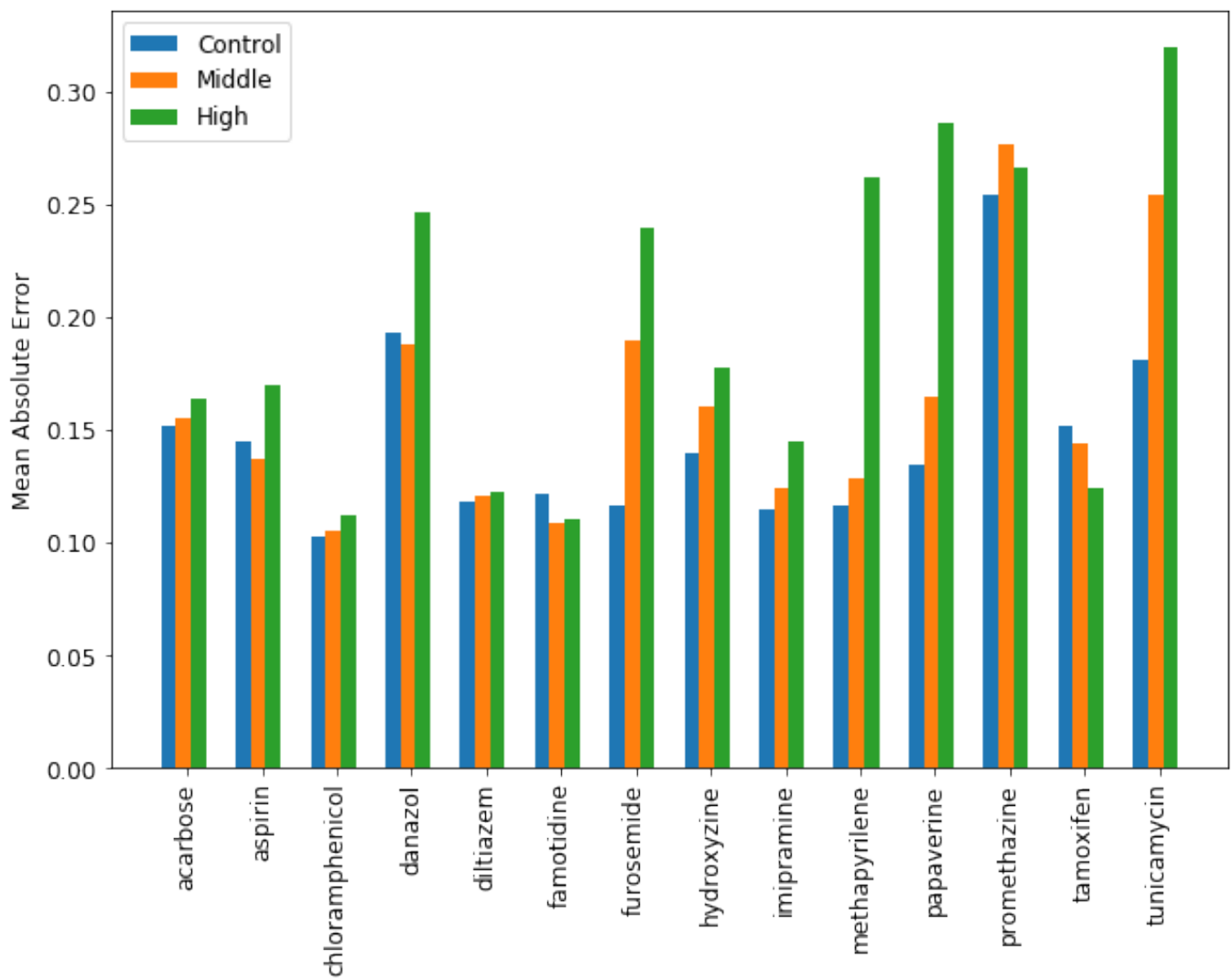

Figure 5. Mean absolute error in predicted expression level versus compound and dose. The values shown are for the 24 hour time points. Means are taken over all genes for each sample.

First, some compounds display a consistently low error level and little or no variation with dose. Examples include chloramphenicol, diltiazem, famotidine, imipramine, and tamoxifen. All of these are compounds for which the majority of the top 100 differentially expressed genes were identical between the true and predicted data.

Second, some compounds display a clear dose effect. Examples include furosemide, methapyrilene, and papaverine. The control samples for these compounds have similar error to most other control samples, but the error increases sharply with dose.  A possible explanation is that these compounds produce changes in gene expression that are significantly different from most of the training compounds, leading to less accurate (though still meaningful) predictions.

Finally, there are a few cases where even the control samples have unusually high error: promethazine, and to a lesser extent danazol and tunicamycin.  In principle all control samples should be interchangeable.  The fact that they are not suggests some other source of variation in the data unrelated to the particular compound being tested.  It is difficult to guess what factor or factors might be responsible.  This serves as an important reminder that any machine learning model can only be as good as the data it is trained with and tested on.

**Saliency Mapping**

For some purposes a neural network can be treated as a black box that produces numbers.  In other cases, it is useful to understand something about how it works.  What input features led to a particular prediction?

With a linear model, we could answer this question simply by looking at the weights.  Every output value would be a linear combination of input values, so we could directly read off how much each one contributed.  A neural network is a nonlinear model, so this is not possible, but we can still form a linear approximation to it.  Given a particular input sample, we compute the derivative of each output with respect to each input.  In the field of computer vision, this technique is known as saliency mapping [7].  For a given output gene, we refer to the absolute value of its derivative with respect to each input gene as the input gene's saliency.  It tells us how sensitive the output is to the level of each rat gene.

We might hope to find that each human gene depends only on a few rat genes, primarily on orthologs.  This would make the model easy to interpret and vindicate the traditional approach of performing translation based on homology.  In practice this is not what we find.  Instead, the predicted expression level for each human gene is the sum of small contributions from a large number of rat genes.  On average, the saliency of the most important rat gene is only about twice that of the 50th most important one.  Each output prediction is weakly influenced by dozens of input values.

This result is not surprising.  It is well known that expression levels for different genes are highly correlated.  For example, it has been shown that given the expression levels of only about 1000 carefully chosen "landmark genes", it is possible to predict the expression levels of the remaining ~21,000 genes in the human genome [8].  Given many highly correlated input features to work with, one should expect the training procedure to select a model that averages over many of them, since this will produce much less noise than using only a few of them.

# Methods

To create the training and test datasets, we began with all *in vitro* human and rat liver samples from Open TG-GATES. RMA normalization [9] was performed with the pyAffy library [10], using the Rat2302_Rn_ENTREZG_22.0.0 and HGU133Plus2_Hs_ENTREZG_22.0.0 CDF files from the BrainArray website [11]. These are custom GeneChip array definition files that correct a variety of problems found in the original definitions created by Affymetrix. They define a total of 14,075 rat genes and 20,352 human genes, each corresponding to a unique Entrez Gene ID [12].

Rat and human arrays were matched up based on compound, dose, and time point. The normalized expression values for the two replicate samples for each condition were averaged. This produced a total of 1122 pairs of corresponding rat and human expression profiles. These were split by compound into training and test sets as described previously.

The mean expression level of each gene over the training set was subtracted from all the data (both training and test sets). This is simply a constant shift which in principle the neural network could learn on its own. In practice, we found that centering the data around zero made the learning problem easier and improved the results.

Our model is a fully connected neural network with one hidden layer of width 20,000 and rectified linear unit activation [13]. It was trained for 1000 epochs with an Adam optimizer [14] and a batch size of 100. The learning rate was initially set to 0.0001, then decayed by multiplying it by 0.9 every 200 steps. 50% dropout was used for regularization [15].

Uncertainty estimation was implemented as described in [16]. This involves two distinct types of uncertainty that are calculated in different ways, then combined to produce a single estimate of the total uncertainty.

- Aleatoric uncertainty refers to the fact that the model does not perfectly fit the training data. It is computed by having the model output an uncertainty estimate for every value it predicts, and using an appropriate loss function to encourage accurate estimates.
- Epistemic uncertainty refers to the fact that multiple models can fit the training data equally well, but they produce different predictions for test data. It is computed by performing prediction many times (50 in this work) with different dropout masks, then computing the variation in output.

The model was implemented with DeepChem 1.2 [17] and TensorFlow 1.9 [18]. Source code for building and training the model is included in the supporting information.

## Discussion

This study is a proof of concept that a neural network can predict gene expression levels in one species based on experimental data from a different species. In this case, we predict *in vitro* human expression based on *in vitro* rat expression. The method should be equally applicable for any pair of species, and also for *in vivo* data. For example, it could be used to predict the clinical effect of a drug candidate based on experimental data from human cell lines or organoids. One simply needs a collection of paired expression profiles to train it on, and the model will learn whatever predictive information is present in the training data. It also could be used to predict the effects of other types of factors: environmental conditions, mutations, etc.

The accuracy of the predicted expression levels varies between samples, but in many cases it is excellent. For the majority of the held out compounds used as a test set, the list of differentially expressed genes identified from the predicted data is very similar to the list identified from actual experimental data. For other compounds the accuracy is lower, but even then the overlap in differentially expressed genes is much greater than would be expected by chance, showing that the model is still producing meaningful information.

As with any machine learning model, the size and quality of the training dataset is critical. A neural network is essentially a way of performing interpolation and extrapolation from existing data. The further it needs to extrapolate, the less accurate its predictions are likely to be. For this study we use the Open TG-GATES database, which contains *in vitro* rat and human data, as well as *in vivo* rat data. Ideally we would like to predict the clinical effects of a drug candidate based on preclinical data, but since Open TG-GATES lacks *in vivo* human data, we were unable to test that. For practical applications of this method, the identification or creation of suitable training datasets is the primary challenge that must be overcome.